\documentclass[a4paper,fleqn]{cas-sc}
\usepackage[numbers]{natbib}
\def\tsc#1{\csdef{#1}{\textsc{\lowercase{#1}}\xspace}}
\tsc{WGM}
\tsc{QE}
\tsc{EP}
\tsc{PMS}
\tsc{BEC}
\tsc{DE}
\begin{document}
\let\WriteBookmarks\relax
\def\floatpagepagefraction{1}
\def\textpagefraction{.001}
\shorttitle{Novel transport phenomena in graphene with strong Spin-Orbit-Interactions}
\shortauthors{T Wakamura et~al.}
%\begin{frontmatter}

\title [mode = title]{Novel transport phenomena in graphene induced by strong spin-orbit interaction}

%\author{T. Wakamura}[type=editor,
                       %auid=000,bioid=1,
                        %role=Researcher,
                        %orcid=0000-0001-7511-2910]
%\author[1,2]{T. Wakamura}
\author{T. Wakamura$^{1,2}$, S. Gu\'{e}ron$^1$ and H. Bouchiat$^1$ }
%\cormark[1]
%\ead{taro.wakamura.ka@hco.ntt.co.jp}
%\ead[url]{www.cvr.cc, cvr@sayahna.org}

%\credit{Experiments, Data analysis, Writing}

%\address[1]{Universit\'{e} Paris-Saclay, CNRS, Laboratoire de Physique des Solides, 91405, Orsay, France}
\address{Universit\'{e} Paris-Saclay, CNRS, Laboratoire de Physique des Solides, 91405, Orsay, France$^1$ \\
NTT Basic Research Laboratories, NTT Corporation, 243-0198, Atsugi, Japan$^2$}
%\address[2]{NTT Basic Research Laboratories, 243-0198, Atsugi, Japan}

%\author[1]{S. Gu\'{e}ron}

%\credit{Supervising, Experiments, Writing}

%\author[1]{H. Bouchiat}

\date{}

\begin{abstract}
Graphene is known to have small intrinsic spin-orbit Interaction (SOI). In this review, we demonstrate that SOIs in graphene can be strongly enhanced by proximity effect when graphene is deposited on the top of transition metal dichalcogenides. We discuss the symmetry of the induced SOIs and differences between TMD underlayers in the capacity of inducing strong SOIs in graphene. The strong SOIs contribute to bring novel phenomena to graphene, exemplified by robust supercurrents sustained even under tesla-range magnetic fields.

\end{abstract}


\maketitle

\section{Introduction}
Since the first experimental demonstration in 2004 \cite{geim}, graphene has been one of the most intriguing materials in condensed matter physics. Novel phenomena have been continuously found, which contrast strikingly with the typical two-dimensional (2D)  electron gas in quantum wells, up to the recent discovery of intrinsic superconductivity and electron correlations in twisted bilayer graphene, exploiting interactions between two neighboring layers \cite{cao}. 

Since graphene is a truly 2D crystalline conductor with  exposed surface electrons, electronic transport properties can be easily modulated via interactions with a substrate, molecules, or another 2D material. 

Graphene is intrinsically rich of novel properties. Its unique Dirac band dispersion brings about extremely high mobility and exotic phenomena such as Klein tunneling \cite{Klein} or unconventional quantum Hall effect \cite{QHE}. However, graphene lacks two properties significant in condensed matter physics: Magnetism and spin-orbit interactions (SOIs). Because electronic transport in graphene is dominated by $\pi$ electrons originated from the 2$p$ orbitals of carbon atoms, their itinerant nature prevents graphene from exhibiting magnetism. Likewise, SOIs are also weak because of the small atomic number of carbon. Recent calculations based on the density functional theory (DFT) estimate 24 $\mu$eV as the intrinsic SOI of graphene \cite{gmitra1}.
Since SOIs are key to implement electrical manipulation of spins in the field of spintronics \cite{spintronics}, and also to drive novel topological phenomena \cite{topological}, there has recently been an intense focus to induce strong SOIs in graphene, with the goal of expanding the functionalities of this novel material.
In early studies on inducing strong SOIs in graphene, many theoretical proposals suggested the use of adatom deposition or molecular functionalization \cite{weeks, hu, balakrishnan}. It is natural to think that deposition of elements with strong SOIs may enhance SOIs in graphene. However, because adatoms or molecules also behave as disorder when deposited on graphene, they degrade transport properties, including the mobility. To maintain the advantageous transport properties in graphene, it can also be considered to deposit graphene on top of a substrate with strong SOIs. As mentioned above, due to exposed 2D electrons, electronic transport in graphene is highly affected by the nature of the substrate, as epitomized by the ultrahigh mobility graphene encapsulated in hexagonal boron-nitride (hBN) \cite{Dean, Wang614}. Since magnetism is successfully induced in graphene by an yttrium iron garnet (YIG, Y$_3$Fe$_5$O$_{12}$) substrate \cite{YIG}, a similar strategy may be successful by selecting an appropriate substrate to induce strong SOIs in graphene and detect them through transport measurements.

Among high spin-orbit (SO) materials, transition-metal dichalcogenides (TMDs) are one of the most reasonable candidates. TMDs are 2D materials with the MX$_2$ composition where M is a transition metal and X a chalcogen such as S,Se or Te.  In our studies we used materials with M either Mo or W and X either S or Se and  the heavy element (Mo or W) is sitting on the A sites of a hexagonal lattice in the trigonal-prismatic (2H-) structure. In the monolayer form, the bottom of the conduction bands and the top of the valence bands are located at inequivalent $K$ and $K'$ valleys, similar to graphene, but a relatively large gap between 1 and 2 eV separates these bands. Therefore these TMDs are semiconductors and much more resistive than graphene. Owing to the heavy elements, the intrinsic SOI is as strong as 1 to a few meV for the conduction bands and 100 - 400 meV for the valence bands \cite{xiao}. Because of the broken A-B sublattice symmetry, the intrinsic SOI in TMDs can be described by a so-called Valley-Zeeman (VZ) type hamiltonian near the band edges where the SO field acts as an out-of-plane Zeeman field. Conservation of time-reversal symmetry requires the direction of the effective Zeeman fields to be opposite at  the $K$ and $K'$ valleys. Electronic and optical properties of these 2D-materials have been thoroughly investigated, and it has already been found that the TMDs properties differ in the monolayer and bulk forms, as exemplified by the direct (monolayer) and indirect (more than two layers) band gap transition \cite{splendiani, mak}. Better coupling is realized between two van-der-Waals 2D materials compared to that between a van-der-Waals 2D material and 3D isotropic material because it is easier to obtain an atomically flat surface experimentally. Therefore 2D TMDs should be good candidates as a substrate to induce strong SOIs in graphene. Moreover, monolayer TMDs are better than bulk (thicker) TMDs as we will discuss below.

In this review,  we demonstrate generation of strong SOIs in graphene by TMDs and present novel transport phenomena observed in the hybrid TMD/graphene system, using not only normal metal contacts but also  superconducting contacts. This review is organized as follows: We briefly mention the sample fabrication process and experimental details in Section 2. Section 3 and Section 4 are devoted to the demonstration of strong SOIs induced in graphene by TMDs. We discuss the difference in the amplitude of SOIs induced in graphene by different TMDs and different thicknesses. The nature and symmetry of the induced SOI are also considered. In Section 5, we present data showing supercurrent in graphene/TMD Josephson junctions which is highly robust to a magnetic field. Finally, we summarize the novel properties of graphene/TMD heterostructures and provide future prospects for this hybrid system as a platform to pursue more exotic effects created by strong SOIs. 

\section{Sample fabrication and experimental details}
\begin{figure}
	\centering
		\includegraphics[scale=.6]{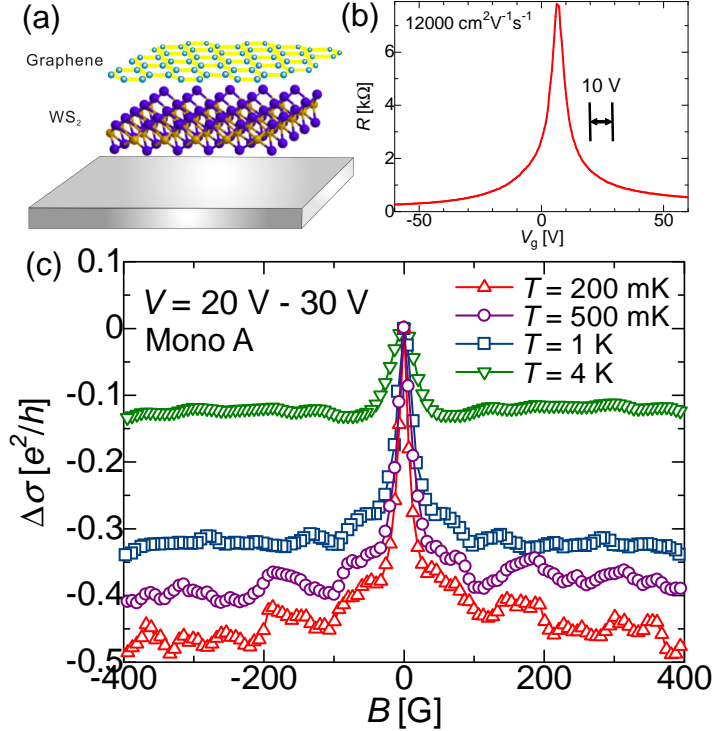}
	\caption{(a) Sketch of the device structure for graphene/monolayer WS$_2$ (Mono WS$_2$) samples is illustrated. Graphene is deposited on the CVD-grown monolayer WS$_2$. (b) Gate voltage ($V_g$) dependence of the resistance ($R$) for Mono WS$_2$ A sample. A clear Dirac peak is observed near $V_g$ = 0, showing that electronic transport in the bilayer is dominated by graphene. The mobility of graphene is relatively high 12000 cm$^2$V$^{-1}$s$^{-1}$. (c) The quantum conductivity correction $\Delta \sigma$ as a function of perpendicular field $B$ from Mono WS$_2$ A sample taken for the electron-doped region at different temperatures. At all temperatures clear WAL peaks are observed, demonstrating that the strong SOI is induced in graphene. The flat tails at higher fields are also a signature of the strong SOI in the system.}
	\label{FIG:1}
\end{figure}

In our studies, we employ three different types of 2D materials: Graphene, TMDs and hBN. Graphene and hBN are always prepared by mechanical exfoliation from graphite and hBN crystals, respectively. The thickness of each flake is determined by the optical contrast on a 285-nm-thick SiO$_2$/Si substrate under the microscope. For TMDs, monolayer WS$_2$ and MoS$_2$ are grown by chemical vapor deposition (CVD) \cite{CVD}, while other TMDs are fabricated by mechanical exfoliation, similarly to graphene and hBN. CVD-grown WS$_2$ is first synthesized on a Si substrate covered with thermally-oxidized SiO$_2$, and then transferred to a fresh SiO$_2$/Si substrate to avoid oxygen vacancies induced in the SiO$_2$ layer during the CVD growth. 

By using these 2D materials prepared separately on SiO$_2$/Si chips, we fabricate heterostructures with polydimethylsiloxane (PDMS) covered with polymethyl methacrylate (PMMA), polypropylene carbonate (PPC) or poly carbonate (PC) \cite{Wang614}. Electrical contacts are formed by electron-beam (EB) lithography, followed by electron-gun evaporation of titanium and gold. For the Josephson junction samples, after the EB lithography the top hBN layer is etched by reactive-ion etching (RIE) and MoRe, a type II superconductor, is sputtered by conventional DC sputtering instead of electron-gun evaporation. 

Low temperature measurements are performed in a dilution refrigerator with a base temperature below 100 mK, and electrical measurements are carried out using a typical lock-in amplifier operating at 77 Hz frequency and excitation current 100 nA unless otherwise specified. 

\begin{figure}
	\centering
		\includegraphics[scale=.8]{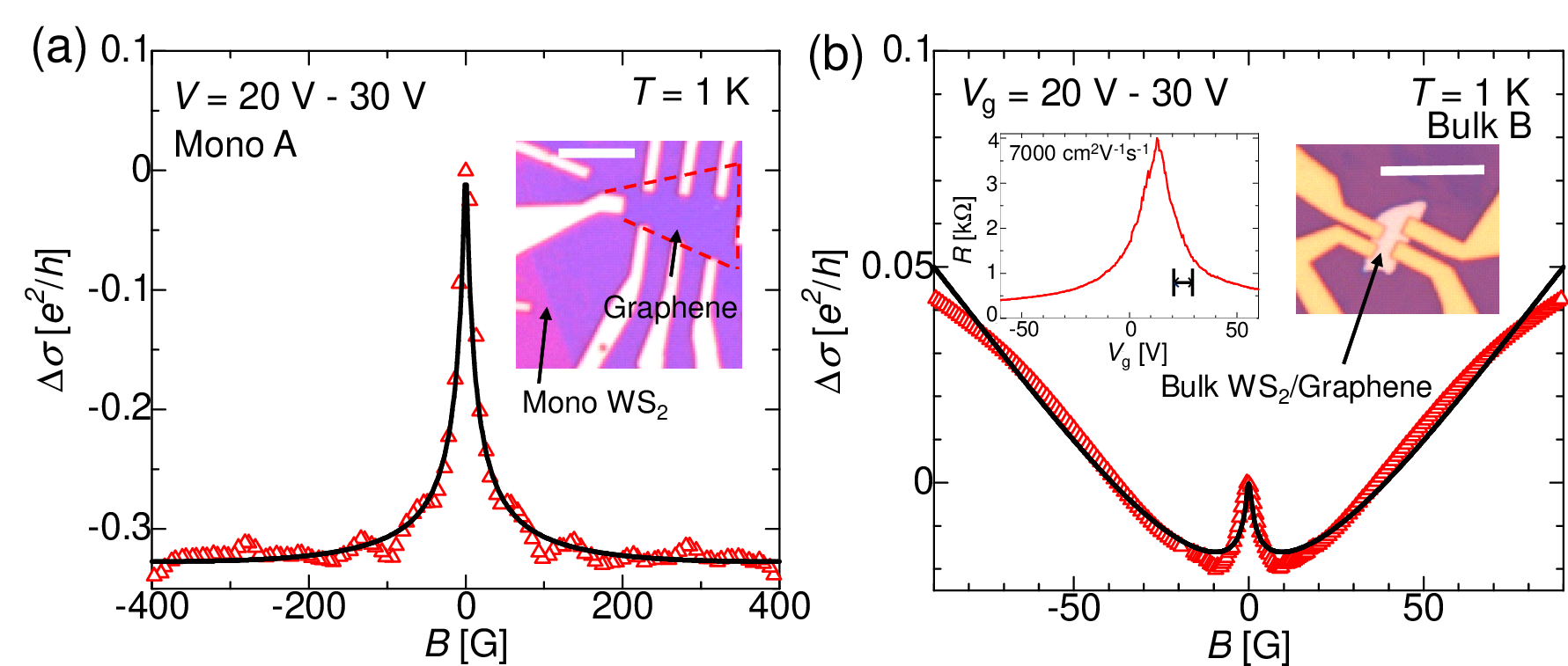}
	\caption{(a) $\Delta \sigma (B)$ from Mono WS$_2$ A sample. The theoretical fit (solid line) reproduces well the experimental result (triangles). The inset shows an optical microscope image of the device. (b) $\Delta \sigma (B)$ from Bulk WS$_2$ B sample in the same $V_g$ region as that in (a). There are striking upturns at higher $B$, showing the dominant WL contribution in these regions. The left inset displays the relation between $R$ and $V_g$ from the same sample, and the right inset shows an optical microscope image of the sample. }
	\label{FIG:2}
\end{figure}
\section{Strong spin-orbit interaction in graphene induced by monolayer and bulk WS$_2$}

We first show evidence of strong SOIs induced in graphene by a CVD-grown WS$_2$ monolayer. Among different TMDs, WS$_2$ has stronger SOIs compared with Mo-based TMDs \cite{xiao}. To induce strong SOIs in graphene by WS$_2$, we fabricate heterostructures as schematically shown in Fig. 1(a). In this structure, monolayer graphene is deposited on monolayer CVD-grown WS$_2$. 

Since WS$_2$ is a semiconductor and much more resistive than graphene, electronic transport is dominated by graphene in the heterostructures especially at low temperatures. Figure 1(b) displays the resistance as a function of the gate voltage ($V_g$) obtained from graphene on monolayer WS$_2$. The sharp Dirac peak is observed between 0 V and 10 V, demonstrating that the transport is dominated by graphene. 

For evaluating the induced SOI in graphene, we employ magnetotransport measurements with magnetic ﬁelds perpendicular to the graphene plane. At low temperatures, the phase coherence of the electron's wave function increases, leading to quantum corrections to the classical Drude conductance for a diffusive conductor. In particular, constructive interferences along time-reversed loops give rise to a low temperature decrease of the conductance. This phenomenon is known as  weak localization \cite{bergmann}. % Considering  a closed  trajectory of an electron  with a diffusive  motion, there is a time-reversed trajectory of the original one when the time-reversal symmetry is conserved. %If the electron's wave functions are coherent during the itinerary through these trajectories, the two electron's wave functions interfere constructively. As a result, these electrons are localized so that the resistance of the system is increased in comparison with the classical (Drude) limit, where the wave nature of electrons is not accounted for. This situation is called weak localization (WL). 
This weak localization correction is canceled in a magnetic field which breaks time-reversal symmetry. Therefore one can measure a resistance decrease with magnetic field, over a characteristic scale corresponding to a flux quantum through the square of the phase coherence length. When the system has strong SOIs, however, the situation is different. SOIs give rise to an additional $\pi$ phase (Berry phase) due to the spin rotation of the electronic wave-functions along time-reversed loops. Therefore wave functions interfere destructively, yielding a positive quantum correction to the conductance in zero field and a positive magnetoresistance. This is called weak antilocalization (WAL). Thus by analyzing magnetoresistance data at low temperatures around the zero magnetic field, we can estimate the amplitude of SOIs in the system based on the weak (anti)localization theory \cite{bergmann}.

We note that WL in graphene is different from that of the typical 2D electron gas. Because of the valley degree of freedom in graphene, graphene has an additional degree of freedom, chirality. Due to the chirality, if one considers a closed trajectory of an electron localized at one valley in the momentum space, the electron acquires a $\pi$ phase during the itinerary even without SOIs, thus naturally exhibit WAL. This is the origin of the absence of localization for Dirac fermions as observed in topological insulators \cite{nomura, liu}, but for real experimental samples of graphene measured at low temperatures, we usually measure WL rather than WAL \cite{tikhonenko}. This is because intervalley scatterings due to disorder in real samples easily break the conservation of chirality and suppress WAL. Therefore WAL in graphene is driven solely by SOIs at low temperatures.  

\begin{figure}
	\centering
		\includegraphics[scale=.75]{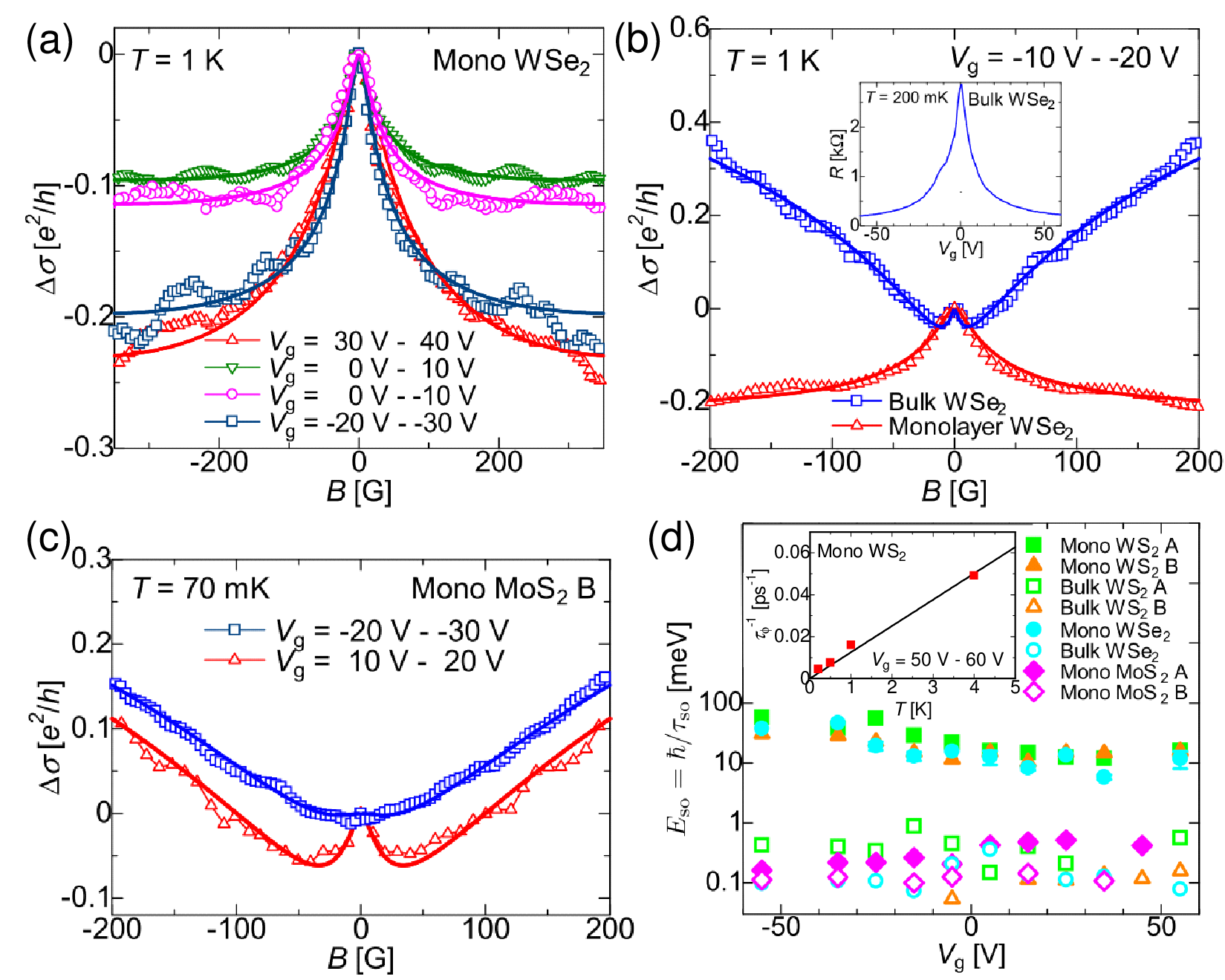}
	\caption{(a) $\Delta \sigma (B)$ from Mono WSe$_2$ taken for the different $V_g$ regions at 1 K. For the all $V_g$ regions clear WAL peaks are observed. (b) Comparison of WAL curves from Mono WSe$_2$ and Bulk WSe$_2$ in the same $V_g$ window. The inset shows the relation between $R$ and $V_g$ from the Bulk WSe2 sample. (c) Similar $\Delta \sigma (B)$ curves from Mono MoS$_2$ both in the electron-doped and hole-doped regions. (d) The spin-orbit energy ($E_{\rm so}$) as a function of $V_g$ obtained from different samples for comparison. It is clear that graphene on monolayer tungsten-based TMDs exhibit the highest SOIs. The inset displays $\tau_\phi^{-1}$ as a function of $T$, which clearly shows that $\tau_\phi^{-1}$ is linearly proportional to $T$, as expected when phase coherence is limited by electron-electron scattering.}
	\label{FIG:2}
\end{figure}

Figure 1(c) shows the quantum conductivity correction ($\Delta \sigma (B) = \sigma (B) - \sigma (0)$) as a function of a perpendicular magnetic field ($B$) taken from a sample with graphene/monolayer WS$_2$ (Mono A) at different temperatures in the electron-doped regime. To suppress the effect of universal conductance ﬂuctuation (UCF), whose amplitude is of the same order of magnitude of $\Delta \sigma (B)$ ($\sim e^2/h$), we average 50 curves with different values of gate voltage in a 10 V window around a given $V_g$. A sharp peak is observed at $B$ = 0 and $\Delta \sigma (B)$ decreases with $B$ at all temperatures, indicating that the system is in the WAL regime. In previous reports on magnetotransport for pristine graphene, $\Delta \sigma (B)$ exhibits WL at low temperatures, thus our results of the observation of WAL demonstrate that the strong SOI is induced in graphene on monolayer WS$_2$. We also fabricate and measure graphene/bulk WS$_2$ (Bulk WS$_2$) samples. Figure 2(a) and (b) are a direct comparison between a Mono WS$_2$ and Bulk WS$_2$ sample in the similar electron-doped region. Both of them exhibit a peak around $B$ = 0, a signature of WAL. Interestingly, we also find that magnetoconductivity curves at higher $B$ for Mono WS$_2$ becomes extremely flat. The field dependence in this region is determined by the competition between the WL and WAL effects with a characteristic upturn if WL is dominant. Thus the weak $B$ dependence at large field indicates that the SOIs induced in graphene by a monolayer WS$_2$ is strong, whereas it is modererate for the case of a WS$_2$ bulk where a sizable upturn is observed.

To evaluate the amplitude of the SOIs precisely, we employ the theoretical formula derived for weak antilocalization in graphene \cite{mccann}. In this formula, contributions from SOIs enter via relaxation times, $\tau_{\rm sym}^{-1} $ and $ \tau_{\rm asy}^{-1}$, where sym (asy) denotes the symmetric (asymmetric) contribution to the SO scattering time with respect to the mirror reflection on the graphene plane. 

The quantum conductivity correction $\Delta \sigma (B)$ is expressed as:
%\begin{multline}
\begin{equation}
\Delta \sigma (B) = -\frac{e^2}{2 \pi h} \left[F \left( \frac{\tau_B^{-1}}{\tau_\phi^{-1}} \right) - F \left( \frac{\tau_B^{-1}}{\tau_\phi^{-1} + 2 \tau_{\rm asy}^{-1}} \right)  -2 F \left( \frac{\tau_B^{-1}}{\tau_\phi^{-1} + \tau_{\rm so}^{-1}} \right) \right],
\end{equation}
%\end{multline}        
where $F(x) = \ln(x)+\psi(1/2 + 1/x)$, with $\psi(x)$ the digamma function and  $\tau_{\rm so}^{-1} = \tau_{\rm sym}^{-1} + \tau_{\rm asy}^{-1}$. $\tau_\phi$ is the phase relaxation time, and $\tau_B = \hbar/4eDB$. The fits yield three parameters $\tau_\phi$, $\tau_{\rm asy}$ and $\tau_{\rm so}$. $\tau_{\rm so}^{-1}$ determines the total amplitude of SOI in the system, whereas from the ratio  $\tau_{\rm asy}/\tau_{\rm sym}$ one can evaluate the symmetry type of SOI.

As shown in Fig. 2(a) as an example, all experimental data are reproduced well by using the equation (1). When focused on the total amplitude of the SOI, these fits provide $\tau_{\rm so}$ = 0.05 ps, independent of temperature. We note that $\tau_\phi$ = 43 ps at 1 K. If we simply define the spin-orbit energy as $E_{\rm so} = \hbar/\tau_{\rm so}$, we obtain $E_{\rm so}$ = 13 meV. This value is three orders of magnitude larger than that of pristine graphene \cite{gmitra1}. We note that due to the flat tail of $\Delta \sigma (B)$ at high fields, $\tau_{\rm so} <$ 0.05 ps provides almost the same curve. Namely, the estimated value of $E_{\rm so}$ shown above is a lower bound of $E_{\rm so}$. 

From the difference in the slope of $\Delta \sigma (B)$ at high fields in Fig. 2(a) and 2(b), we can deduce that the induced SOIs seem smaller in Bulk WS$_2$ than Mono WS$_2$, as already mentioned. To estimate the amplitudes of SOIs in Bulk WS$_2$ more accurately, we also fit the experimental data taken from Bulk WS$_2$ using equation (1). The fits yield  $\tau_\phi$ = 36 ps and $\tau_{\rm so}$ = 4 ps, equivalent to $E_{\rm so}$ = 170 $\mu$eV. This value is still larger than that of pristine graphene, but much smaller than that obtained by using monolayer WS$_2$. Therefore monolayer WS$_2$ can induce stronger SOIs in graphene than bulk WS$_2$. The difference in the efficiency to induce strong SOIs between monolayer and bulk TMDs will be discussed in detail in the next section.

\section{SOIs in graphene induced by different TMDs with different thickness}

To confirm that TMDs can universally induce strong SOIs in graphene, we measure graphene/TMD heterostructures with different types of TMDs with different thickness. In our study we focus on semiconducting 2$H$- structure TMDs, specifically WS$_2$, WSe$_2$ and MoS$_2$. As already shown in the previous section, WS$_2$ can induce strong SOIs in graphene through the proximity effect. Therefore it is also interesting to use WSe$_2$ to induce strong SOIs in graphene since tungsten-based TMDs such as WS$_2$ and WSe$_2$ are reported to have stronger SOI than molybdenum-based SOIs, and also WSe$_2$ has stronger intrinsic SOI than  WS$_2$ \cite{xiao}. Recent theoretical calculations reveal that for the valence band the SOI is 460 meV for monolayer WSe$_2$ while it is 430 meV for monolayer WS$_2$ \cite{xiao}. Therefore stronger SOIs are expected to be induced in graphene by monolayer WSe$_2$.

We fabricate graphene/monolayer WSe$_2$ (Mono WSe$_2$) and graphene/bulk WSe$_2$ (Bulk WSe$_2$) and estimate SOIs induced in graphene via weak (anti)localization measurements. Figure 3(a) shows $\Delta \sigma (B)$ obtained from Mono WSe$_2$ at different gate voltages. All curves exihibit WAL peaks, demonstrating that strong SOIs are induced in graphene. Similar to the graphene/WS$_2$ samples, we can also compare the amplitudes of the SOIs induced in graphene by monolayer and bulk WSe$_2$ flakes. Figure 3(b) is a direct comparison between the two curves from the Mono WSe$_2$ and Bulk WSe$_2$ samples. While there is a striking upturn at high fields for Bulk WSe$_2$, a higher peak around $B$ = 0 and a flat tail are observed for Mono WSe$_2$. This is in agreement with our previous observations on graphene/WS$_2$ structures and confirms that monolayer TMDs can induce stronger SOIs in graphene than bulk TMDs can. 

We also expand our study to evaluate SOIs induced in graphene by molybdenum-based TMD (MoS$_2$). Because the atomic number of molybdenum is smaller than that of tungsten, its intrinsic SOI is also smaller and recent theoretical calculations reported 150 meV for the value of the intrinsic SOI in the valence bands of monolayer MoS$_2$, almost three times smaller than those of the tungsten-based TMDs \cite{xiao}. Nevertheless, MoS$_2$ is still usable to enhance SOIs in graphene. In Fig. 3(c) $\Delta \sigma(B)$ from graphene/monolayer MoS$_2$ (Mono MoS$_2$) is plotted both for the electron-doped and hole-doped regions. Especially in the electron-doped region, a clear WAL peak is visible around $B$ = 0, a signature of the strong SOIs in graphene. By contrast, the peak in the hole-doped region is weak.

$E_{\rm so}$ for graphene with different TMDs with different thickness is summarized in Fig. 3(d) for different gate voltages. It reveals that monolayer tungsten-based TMDs (WS$_2$ and WSe$_2$) can induce the strongest SOIs in graphene, and that their bulk counterparts induce weaker SOIs.  We find that the amplitude of the SOIs induced by a monolayer MoS$_2$ is comparable to those induced by bulk WS$_2$ and WSe$_2$. Also, the gate dependence of the induced SOI is found to be rather weak. We note that in the hole-doped region for Mono WS$_2$ and Mono WSe$_2$ samples, temperature-independent background signals are superimposed on the WAL signals. Since WL and WAL are quantum effects and temperature dependent, we subtract those temperature-independent backgrounds when we carry out fits based on equation (1) \cite{wakamura1, wakamura2}.

While it seems reasonable that the SOIs generated by tungsten-based TMDs should be stronger than those induced by molybdenum-based TMDs,  the difference between monolayer and bulk TMDs is more difficult to understand in terms of the capacity to induce SOIs in graphene . 

There are several differences between monolayer and bulk TMDs. For instance, TMDs such as WS$_2$, WSe$_2$ and MoS$_2$ are in the trigonal-prismatic 2$H$- structure. In the monolayer form, this structure is inversion asymmetric, but in the bulk form, it is inversion symmetric since two neighboring layers are coupled by inversion symmetry. This difference with respect to space inversion operation between monolayer and bulk TMDs is reflected in the band structure of each system. For monolayer TMDs, the bands are spin-split due to the inversion symmetry breaking, whereas they are spin degenerate for bulk TMDs \cite{xiao, TMDbands1, TMDbands2}. Recent theoretical studies point out that the strength of the spin splitting in the band structure severely affects the amplitude of SOIs induced in graphene \cite{koshino}. Therefore monolayer TMDs, which have spin-split bands, are superior to bulk TMDs to generate stronger SOIs in graphene.

Another factor to consider is the possible interface mismatch between bulk TMD crystals and graphene. As reported previously, graphene in real samples is not ideally flat and includes ripples and bubbles. Monolayer TMDs are more flexible so that they can conform to the neighboring graphene layer, but it is more difficult in the case of bulk TMDs. This interface mismatch may reduce the interaction potential between conduction electrons in graphene and atomic orbitals in the TMD. There are no supporting theoretical \textit{scenarii} for this difference at the time of writing, thus further experimental and theoretical studies are required. 

\section{Identification of the dominant type of SOIs}
\begin{figure}
	\centering
		\includegraphics[scale=.9]{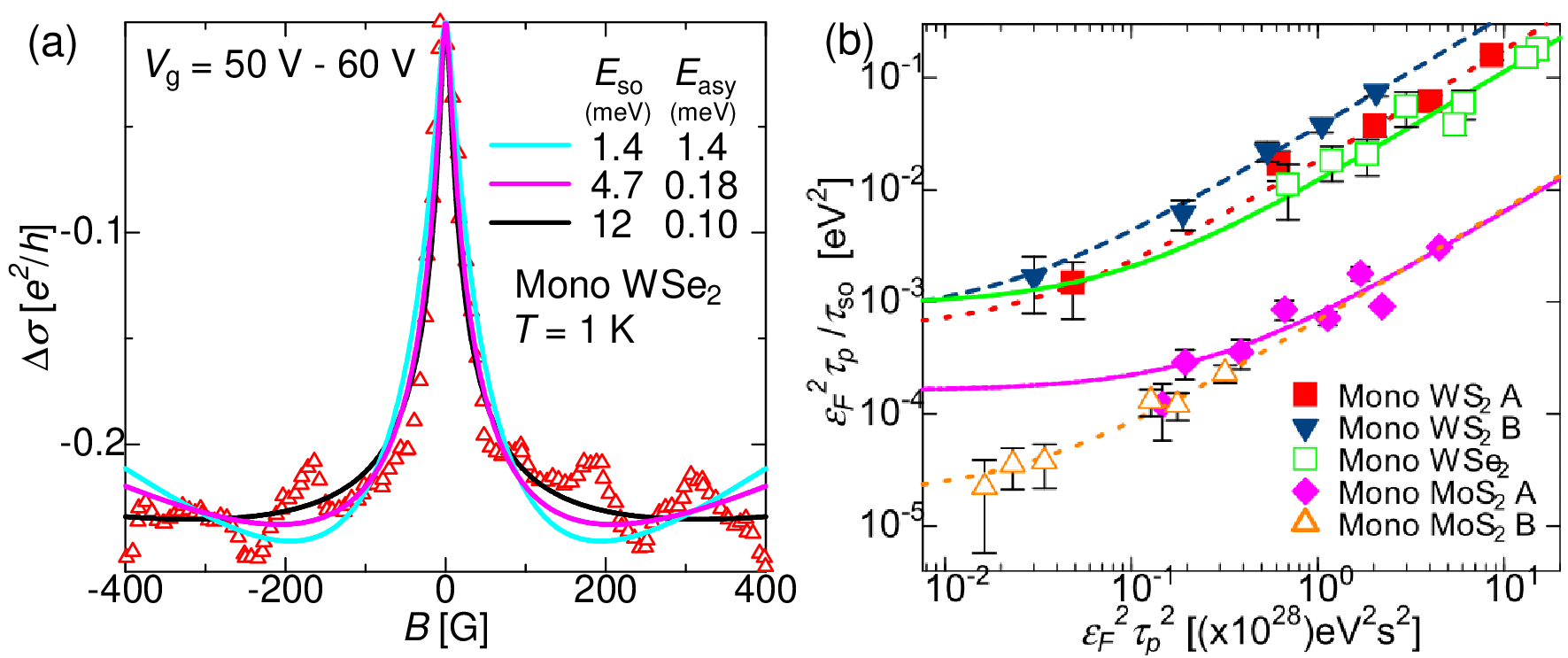}
	\caption{(a) $\Delta \sigma (B)$ from Mono WSe$_2$ sample. The best fit (black solid line) is obtained for $\tau_{\rm sym} \ll \tau_{\rm asy}$, while other fits calculated intentionally with different ratio between $\tau_{\rm sym}$ and $\tau_{\rm asy}$ clearly deviate from the experimental results. (b) The relation between $\varepsilon_F^2 \tau_p^2$ and $\varepsilon_F^2 \tau_p/\tau_{\rm so}$. The deviation from the linear relation provide the estimate of the contribution from the EY spin relaxation mechanism.}
	\label{FIG:PRB1}
\end{figure}

In the previous sections, we evaluated the amplitude of SOIs induced in graphene by the TMDs from $\tau_{\rm so}$ obtained from the fits based on (1). %The intrinsic SOI in graphene, called also as the Kane-Mele (KM) SOI, is categorized as the symmetric SOI, %
 Three different types of SOIs are theoretically predicted to be induced in graphene on TMDs; Kane-Mele (KM, or intrinsic), Rashba and Valley-Zeeman (VZ) SOI. KM SOI is the intrinsic SOI, thus it exists even in pristine graphene. Rashba SOI is allowed in a system with broken inversion symmetry, especially when the $z \rightarrow -z$ symmetry is broken in the case of 2D materials. Therefore large Rashba SOI may be expected in monolayer graphene on TMD intuitively, but surprisingly it is almost negligible as we will discuss below. In contrast to monolayer graphene, Rashba SOI may be dominant in the case of bilayer graphene on TMD as reported in \cite{wang2}. On the other hand, VZ SOI breaks A-B sublattice symmetry, and it becomes significant when heavy elements (such as Mo or W) contribute to the sublattice symmetry breaking \cite{fabianprb}. By evaluating the ratio between $\tau_{\rm sym}$ and $\tau_{\rm asy}$ ($\tau_{\rm asy}/\tau_{\rm sym}$) in the fits based on the equation (1), we can determine the symmetry of the SOI principally induced in graphene by the TMDs. For all investigated samples, this ratio is found to be larger than 10, which indicates that the Rashba component of the SOIs acting on the in-plane spin components leading to the $\tau_{\rm asy}$ scattering time is much weaker  than the out-of-plane components, which determine the $\tau_{\rm sym}$ scattering time. For graphene on TMDs, two types of $z \rightarrow -z$ symmetric SOI may be considered, the intrinsic KM SOI and VZ SOI. The former and the latter are respectively symmetric and asymmetric about the A-B symmetry of the hexagonal lattice. We note that the lattice parameters of TMDs and graphene are different.  Moreover, we do not specifically align the relative orientations of the two layers. Therefore it is \textit{a priori} tempting to assume that carriers in graphene are scattered randomly by heavy TMD atoms. %In that case, the KM SOI should be predominant. I think you found  from numerics that this is not true, there is no  induced KM because the avrege of lambda_AA is zero... 
 However, it is shown theoretically \cite{koshino, david} that this is not an appropriate picture  and that in spite of the incommensurablity between the two lattices , the TMD/graphene bilayer retains a $C_3$ symmetry  whereas the initial $C_6$ of Graphene is lost.  Actually a VZ SOI is induced in graphene with an amplitude of a few meV depending on the nature of the TMD.  In the following we show that our experimental data are consistent with these findings.
% ***** still need to agree on a sentence here****

Figure \ref{FIG:PRB1}(a) shows fits for the Mono WSe$_2$ sample with different ratios between $\tau_{\rm sym}$ and $\tau_{\rm asy}$. The fits with $\tau_{\rm sym} \ll \tau_{\rm asy}$ unambiguously give  a much better description of the experimental results. This demonstrates that the symmetry of the SOIs induced in graphene by the TMDs is predominantly $z \rightarrow -z$ symmetric.

\begin{table}[width=.9\linewidth,cols=3,pos=htb]
\caption{$\Delta_{\rm EY}$ and $\Delta_{\rm DP}$ obtained from the fits using (\ref{eq_spin_relax}) for each sample}\label{tbl1}
\begin{tabular*}{\tblwidth}{@{} LLL@{} }
\toprule
Sample & $\Delta_{\rm EY}$ [meV] & $\Delta_{\rm DP}$ [meV]\\
\midrule
Mono WSe$_2$ & 48.0 $\pm$ 12.2 & 3.3 $\pm$ 0.10\\
Mono WS$_2$ A & 27.4 $\pm$ 2.8 & 4.4 $\pm$ 0.047\\
Mono WS$_2$ B & 32.9 $\pm$ 3.8 & 6.4 $\pm$ 0.14\\
Mono MoS$_2$ A & 10.3 $\pm$ 1.7 & 0.87 $\pm$ 0.035\\
Mono MoS$_2$ B & 4.3 $\pm$ 0.035 & 0.86 $\pm$ 3.5$\times 10^{-3}$\\
Bulk WSe$_2$ & 11.6 $\pm$ 2.1 & 0.38 $\pm$ 0.013\\
Bulk WS$_2$ A & 8.9 $\pm$ 3.7 & 0.73 $\pm$ 0.028\\
Bulk WS$_2$ B & - & 0.72 $\pm$ 0.022\\
\bottomrule
\end{tabular*}
\end{table}

  We  first note that the KM and VZ SOI, both of which are $z \rightarrow -z$ symmetric, give rise to different spin relaxation mechanisms in graphene. There are two spin relaxation mechanisms in graphene, Elliot-Yafet (EY) and D’yakonov-P\'{e}rel (DP). These two mechanisms provide different dependences of $\tau_{\rm so}$ on $\tau_p$, where $\tau_p$ is the momentum relaxation time. Whereas $\tau_{\rm so} \propto \tau_p$ for the EY mechanism, $\tau_{\rm so} \propto \tau_p^{-1}$ for the DP. The KM SOI contributes to the EY spin relaxation mechanism since the DP mechanism requires spin-splitting due to inversion symmetry breaking. On the other hand, since the VZ SOI originates from broken sublattice symmetry, it leads to a DP-type spin relaxation. We neglect the smaller contributions from the Rashba SOI . Each contribution (EY or DP) can be determined by ﬁtting the relation between $\tau_{\rm so}$ and the momentum relaxation time $\tau_p$ following the equation \cite{zomer}:
\begin{equation}
\frac{\varepsilon_F^2 \tau_p}{\tau_{\rm so}} = \Delta_{\rm EY}^2 + \left( \frac{4 \Delta_{\rm DP}^2}{\hbar^2} \right)
\varepsilon_F^2 \tau_p^2
\label{eq_spin_relax}
\end{equation}
where $\Delta_{\rm EY(DP)}$ is the amplitude of spin-orbit coupling leading to the EY (DP) mechanism and $\varepsilon_F$ is the Fermi energy. Figure \ref{FIG:PRB1}(b) displays the relation in equation (\ref{eq_spin_relax}) for each sample. %Finite gradients are due to the contribution from the EY mechanism. 
It is clear from these expressions that the DP scattering mechanism is  dominant in the highly doped region whereas the EY is dominant in the vicinity of the Dirac point.
In Table 1, we show the fitting parameters $\Delta_{\rm EY}$ and $\Delta_{\rm DP}$ obtained from the fits for each device. The DP energy scales found are in good agreement with theoretical predictions concerning the VZ SOI \cite{david}.
%We can figure out that the EY contribution is much more dominant than 
We also find a large EY contribution whose origin is more difficult to explain since the KM intrinsic SO contribution is expected to be small compared to the VZ one, considering the recent theoretical studies \cite{koshino, david}. Up to now we do not have a good explanation for these large values of $\Delta_{\rm EY}$. However we must note that these values are extracted from data taken close to the Dirac point where the formation of a disordered network of electron-hole puddles dominates the properties of graphene \cite{yacoby} and  $\varepsilon_F$ is ill-defined, resulting in possible important errors in the determination of $\Delta_{\rm EY}$ from (\ref{eq_spin_relax}).

\section{Superconducting Josephson junctions with graphene/WS$_2$ heterostructures}

Now that we have confirmed that strong SOIs can be induced in graphene by the TMDs, it is also tempting to integrate graphene/TMD heterostructures into superconducting Josephson junctions. There have already been many studies on Josephson junctions with graphene \cite{Amet, borzenets, benshalom, calado, Zhao}, but transport properties of supercurrent in graphene with strong SOIs have never been explored. SOIs are an essential ingredient to realize topological superconductivity, thus it is of great significance to investigate effects of strong SOIs on supercurrent transport in graphene. In this section, we present robust supercurrent observed in Josephson junctions with graphene/WS$_2$ heterostructures.

\begin{figure}
	\centering
		\includegraphics[scale=.9]{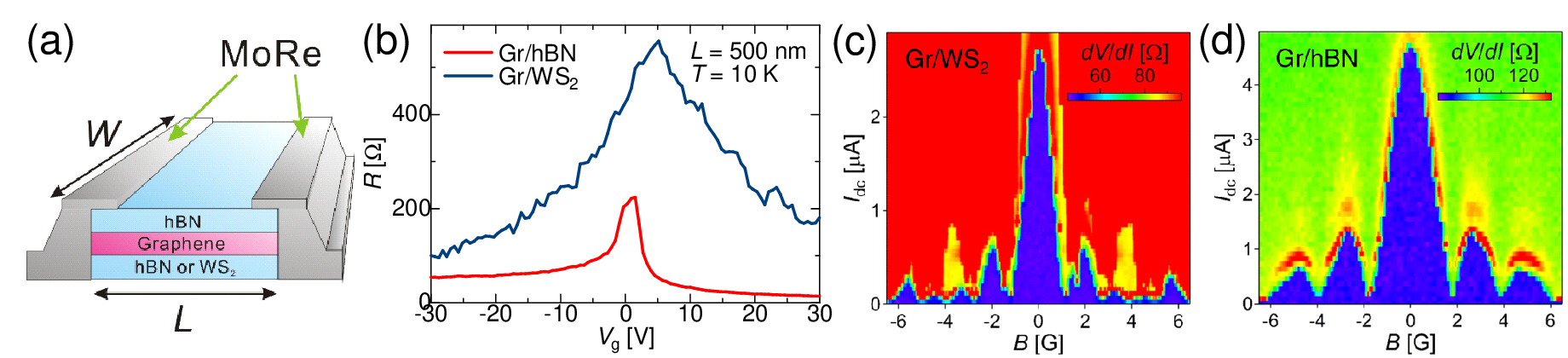}
	\caption{(a) Schematic illustration of superconducting Josephson junction composed of hBN/graphene/WS$_2$ or hBN/graphene/hBN in contact with MoRe superconducting electrodes. (b) Gate voltage dependence of resistance for Gr/hBN and Gr/WS$_2$ junctions. (c) $dV/dI$ plotted for both the dc current ($I_{\rm dc}$) and field ($B$) for a $L$ = 500 nm Gr/WS$_2$ junction. Fraunhofer-like oscillations are visible. (d) Similar $dV/dI$ plot as in (c) obtained from a $L$ = 500 nm Gr/hBN junction. Larger critical current $I_c$ is observed compared with that for the Gr/WS$_2$ junction in (c).}
	\label{FIG:SC1}
\end{figure}

The device structure is illustrated in Fig. \ref{FIG:SC1}(a). In this study we employ superconducting electrodes made of MoRe because it provides a good superconducting contact to graphene without any buffer layers, and possesses high critical temperature ($T_c \sim$ 10 K) and high critical magnetic field ($H_{c2} \sim$ 8 T). The MoRe electrodes are connected to two types of structures, either hBN/graphene/WS$_2$ (Gr/WS$_2$) or hBN/graphene/hBN (Gr/hBN) heterostrucures, forming a superconducting Josephson junction. The width of graphene is around 10 $\mu$m for all junctions, and the length of the junction ranges from 100 nm to 500 nm, depending on the sample.

We first show the gate dependence of the resistance in Fig. \ref{FIG:SC1}(b). There is a striking difference between Gr/WS$_2$ and Gr/hBN junctions, and the Dirac peak for the Gr/WS$_2$ junction is much broader than that of the Gr/hBN junction. This indicates that the mobility of graphene on WS$_2$ is much  smaller than that on hBN. While a recent study reported that TMDs can be a substrate as good as hBN for graphene \cite{Young}, since our samples employ monolayer CVD-grown WS$_2$, the surface of flakes is not as flat as multilayer hBN, and also there may remain some residues on the surface, originating from the transfer process of CVD-grown WS$_2$ from the original substrate used during the growth to a fresh substrate for measurements (see Section 2). We note that all Gr/WS$_2$ junctions with different $L$ have a lower mobility than the Gr/hBN junctions with the same $L$. For $L$ = 100 nm and 200 nm Gr/hBN junctions, we observe Fabry-P\'{e}rot-like oscillations in the hole doped region (not shown) in the normal state. This is in agreement with previous reports \cite{calado, benshalom}, and demonstrates that these two junctions are in the ballistic regime.

\begin{figure}
	\centering
		\includegraphics[scale=.9]{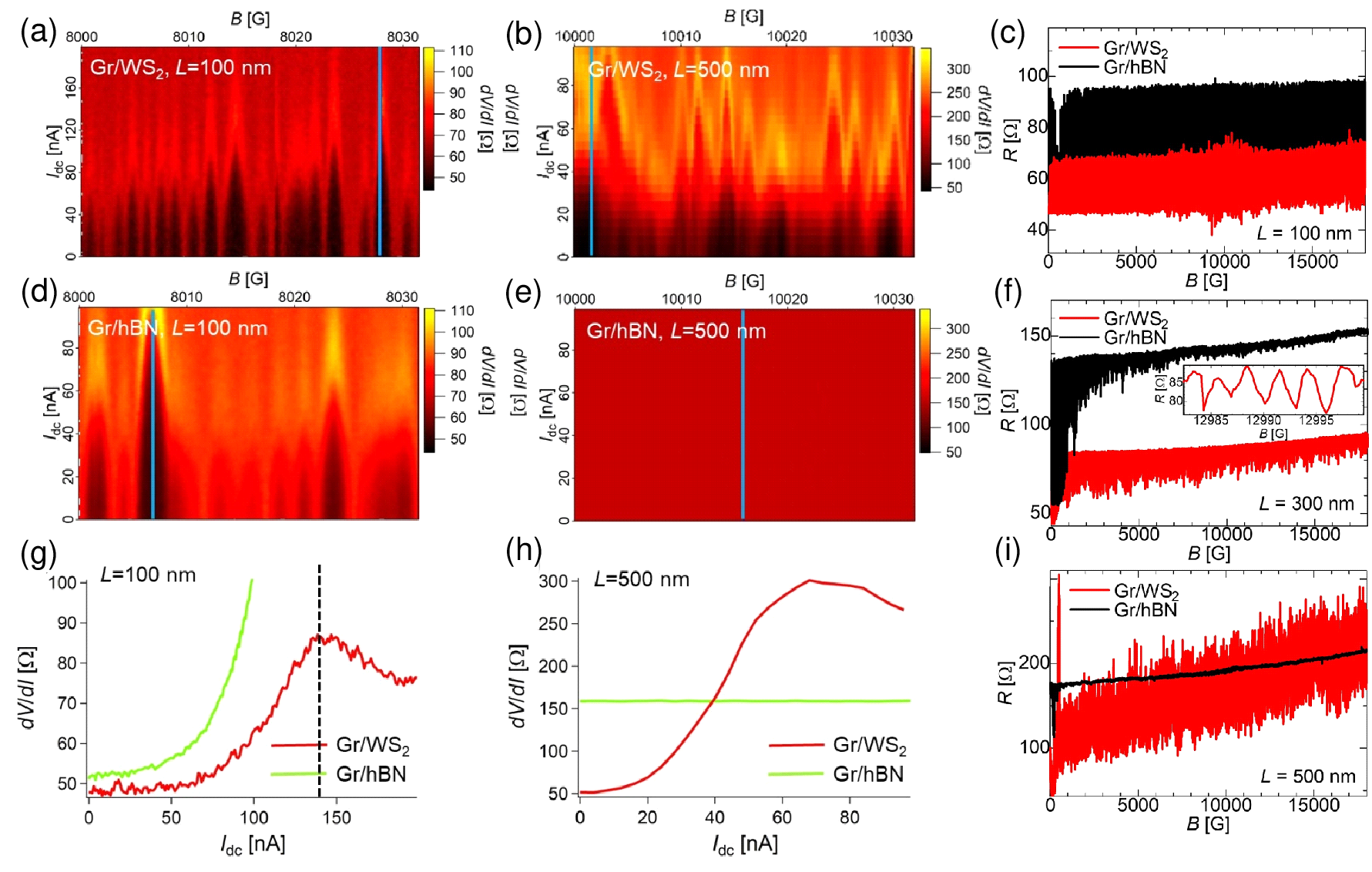}
	\caption{(a) and (d) are $dV/dI$ as a function of both $I_{\rm dc}$ and $B$ around $B$ = 8000 G for $L$ = 100 nm Gr/WS$_2$ ((a)) and Gr/hBN ((d)) junctions. Both junctions show the suppressed $dV/dI$ at certain magnetic fields ((g)), signatures of the induced superconductvity in graphene. (b) and (e) display $dV/dI$ around $B$ = 10000 G for $L$ = 500 nm Gr/WS$_2$ ((b)) and Gr/hBN ((e)) junctions. While the Gr/WS$_2$ junction exhibits the suppressed $dV/dI$, it is competely absent in the Gr/hBN junction ((h)). (c), (f) and (i) show the zero-bias differential resistance $R$ as a function of $B$ for Gr/WS$_2$ and Gr/hBN junctions with different $L$. Clear differences in the decay of the amplitudes of the oscillations are observed between Gr/WS$_2$ and Gr/hBN junctions.}
	\label{FIG:SC2}
\end{figure}
Figure \ref{FIG:SC1}(c) and (d) display the differential resistance ($dV/dI$) as a function of magnetic field ($B$) for $L$ = 500 nm Gr/WS$_2$ (c) and Gr/hBN (d) junctions. To measure $dV/dI$, a small ac current ($I_{\rm ac}$) is added to the dc current ($I_{\rm dc}$). Fraunhofer-like oscillations are clearly observed, a signature of supercurrent flowing through the junctions. Slight deviations from the ideal shape of oscillations may be due to small inhomogeneities of the current distribution in the junction. The Gr/hBN sample exhibits larger critical current ($I_c$) at zero field, consistent with the higher mobility of graphene.

While superconducting properties are better for Gr/hBN junctions than Gr/WS$_2$ around the zero field, these characters become dramatically different at higher fields. Figure \ref{FIG:SC2} displays $dV/dI$ around 1 T (10000 G). For $L$ = 100 nm junctions, even around 8000 G, dips are observed in $dV/dI$ and $I_c$ oscillates with $B$ for both junctions (Fig. \ref{FIG:SC2}(a) and (d)). In the previous report on graphene ballistic Josephson junctions, field-dependent and sample-specific differential resistance dips at a low current were also observed around 5000 G, and the $B$ and $V_g$ regions of low $dV/dI$ were termed 'superconducting pockets' \cite{benshalom}. In our shortest samples, the superconducting pockets are still visible around $B$ = 16000 G for the Gr/WS$_2$ junction (not shown). We note that all Gr/WS$_2$ junctions, even the shortest one with $L$ = 100 nm, are in the diffusive limit due to the shorter mean free path $l_e$, whereas the $L$ = 100 nm Gr/hBN junction is in the ballistic limit.

While signatures of robust supercurrent are observed for both Gr/hBN and Gr/WS$_2$ junctions for $L$ = 100 nm samples, a striking difference emerges for longer junctions ($L$ = 500 nm). Figure \ref{FIG:SC2} (b) and (e) display the colour-coded differential resistance $dV/dI$ as a function of $I$ and $B$, for $B$ around 10000 G for Gr/WS$_2$ ((b)) and Gr/hBN ((e)) junctions. For the Gr/WS$_2$ junction superconducting pockets are still observable. By contrast, they are completely suppressed for Gr/hBN junctions. In Fig. \ref{FIG:SC2} (g) and (h) we compare crosssectional cuts of $dV/dI$ for both samples with $L$=100 nm ((g)) and $L$ = 500 nm ((h)). The suppressed (Gr/WS$_2$) and flat (Gr/hBN) $dV/dI$ are shown more explicitly in Fig \ref{FIG:SC2}(h).  

\begin{figure}
	\centering
		\includegraphics[scale=.9]{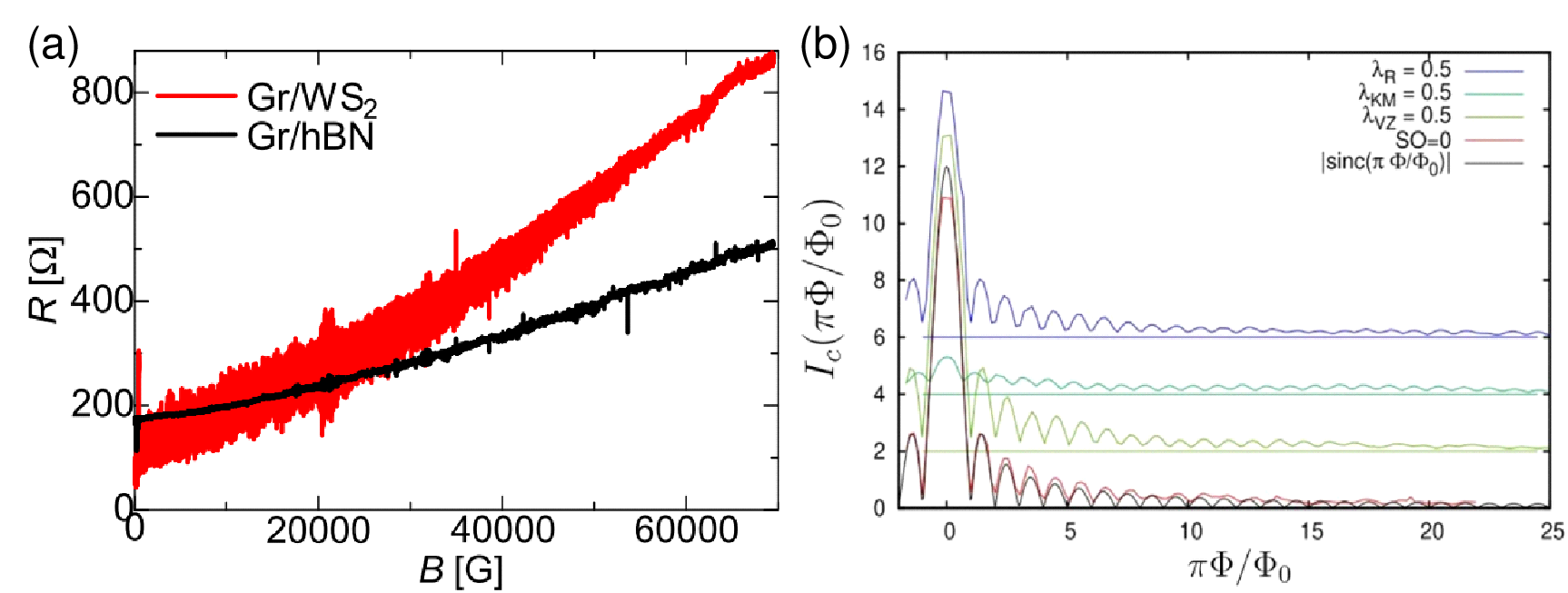}
	\caption{(a) Zero bias differential resistance $dV/dI (V_{\rm dc}=0) \equiv R$ with $B$ from 0 G up to 70000 G from the $L$ = 500 nm Gr/WS$_2$ and Gr/hBN junctions. Oscillations are still visible even at $B$ = 70000 G. (b) Simulated $I_c$ as a function of the normalized flux quantum ($\Phi/\Phi_0, \Phi_0 = h/2e$) with and without SOIs. As for SOIs, three types of SOI are considered: Rashba ($\lambda_{\rm R}$), Kane-Mele ($\lambda_{\rm KM}$) and valley-Zeeman ($\lambda_{\rm VZ}$) SOIs. For the details of the simulation, please refer to Supplemental Material of \cite{wakamura3}.}
	\label{FIG:SC3}
\end{figure}

To investigate the evolution of $dV/dI$ in a broader range of field, we measure the zero bias differential resistance ($dV/dI (I_{\rm dc}=0) \equiv R$) with increasing field. Figure \ref{FIG:SC2} (c), (f) and (i) show the relation between $R$ and $B$ for $L$ = 100 nm ((c)), $L$ = 500 nm ((i)) and also for $L$ = 300 nm ((f))  Gr/WS$_2$ and Gr/hBN junctions. For the $L$ = 100 nm junctions, the amplitudes of oscillations are comparable for both Gr/WS$_2$ and Gr/hBN junctions up to 18000 G. However, for the $L$ = 300 nm junctions, while the Gr/WS$_2$ junction exhibits relatively large amplitude of oscillations even around $B$ = 18000 G, the amplitude is strongly suppressed for the Gr/hBN junction. The difference is much more striking for the $L$ = 500 nm junctions. Whereas the oscillations are rapidly suppressed for $B$ less than 1000 G for the Gr/hBN junction, $R$ continuously oscillates up to 18000 G for the Gr/WS$_2$ junction. We show that the oscillations are still observable even at 70000 G in Fig. \ref{FIG:SC3}(a) \cite{wakamura3}.

Now we discuss the possible origin of superconducting pockets for these junctions. As for $L$ = 100 nm ballistic Gr/hBN junctions, Andreev bound states mediated by chaotic ballistic billiard paths located at the edges of graphene have already been proposed as an origin \cite{benshalom}. Those chaotic paths can be regarded as a ballistic analog of the quasi-classical phase-coherent paths and produce mesoscopic fluctuations of the supercurrent $\delta I_c = \sqrt{\langle I_c^2 \rangle - \langle I_c \rangle^2}$. In the case of ballistic short junctions, $\delta I_c$ is estimated as

\begin{equation}
\delta I_c \sim \frac{e \Delta_0}{\hbar},
\end{equation}
where $\Delta_0$ denotes the superconducting gap at $T$ = 0 \cite{beenakker}. 

This formula yields $\delta I_c \sim$ 240 nA for short Gr/hBN junctions, slightly higher than the experimentally observed value. We note that the above formula is calculated in the zero field limit. For finite magnetic fields, a recent theoretical study for 2D ballistic junctions provides a picture of supercurrent distribution at high fields \cite{Falko}. As $B$ increases, supercurrent is localized close to the edges, and $I_c$ decays faster ($I_c \propto 1/B^2$) than the typical decay of $I_c$ around the zero field ($I_c \propto 1/B$). By contrast, supercurrent fluctuations $\delta I_c$ persists even at high fields, and are expressed as $\delta I_c = \alpha E_T / \Phi_0$ with $\alpha = 2 \pi /9 \sqrt{3}$. This estimate provides $\delta I_c \sim$ 200 nA for the $L$ = 100 nm Gr/hBN ballistic junctions, in qualitative agreement with our experimental results. However, these high field supercurrent fluctuations are specific for ballistic junctions, and not expected for diffusive junctions.

In order to explain the robust supercurrent signatures which we find in the diffusive Gr/WS$_2$ junctions, it thus seems necessary to account for the role of SOIs. 
SOIs favor the formation of edge states, epitomized by the topological quantum spin Hall phase \cite{QSHE}. We note that it is unlikely for our results to be relevant to the topological edge states because our measurements are carried out in the highly-doped region. On the other hand, edge states can also exist in a non-topological system, and coexist with bulk states of the same energy \cite{beenakker2}. Such edge states are generally sensitive to scattering, but some degree of protection against smooth disorder may exist if the edge states are well separated from bulk states in momentum space. In addition, spin can also provide additional protection if the spins of the edge state and those of the nearby bulk band are opposite. In the case of graphene on WS$_2$, the analysis of weak antilocalization experiments, see previous sections and \cite{wakamura1, wakamura2, wang1, wang2, zihlmann}, has shown that the induced SOIs have both a Rashba-type in-plane component and a more than ten-folds larger out-of-plane component, predominantly of valley-Zeeman type probably.
The combined effect of those two types of interactions was demonstrated theoretically to generate non-topological edge states along zigzag edges \cite{frank}. For exploring whether such edge states may explain the persistance and oscillations of supercurrent at high fields, we have carried out simulations of $I_c$ as a function of $B$ in graphene strips containing different types of SOIs (Fig. \ref{FIG:SC3}(b)). We find that  supercurrent persists up to higher fields with SOIs than that without SOIs, even when disorder is included. Due to divergently increasing calculational time with increasing number of the flux quanta, we unfortunatly cannot reach the regime where thousands of flux quanta thread the junction. Further theoretical studies are awaited.

While we carried out our supercurrent experiments in the highly doped region, it may be interesting to contemplate the properties of the edge state from the view point of topological physics. Nontopological edge states are theoretically predicted as shown in \cite{frank}, but no theoretical works address the graphene VZ SOI system proximitized by superconductivity. As evidenced in previous studies, supercurrent transport reveals topological properties of the system \cite{chuan1, chuan2, murani}. Considering the fact that the edge states observed in our study are protected against disorder, novel topological phases are possible. Theoretical works such as Majorana fermions in a half-metallic wire on an Ising superconductor \cite{zhou} may be helpful to obtain some insights to study these novel phases. 

We comment on the relation between those edge paths and the chiral edge states of the quantum Hall (QH) regime. The QH regime develops when the mean free path $l_e \gg 2r_c$, where $r_c$ is the cycrotron radius ($r_c = \hbar k_F / eB$) \cite{Amet}. For example at $V_g$ = 60 V and $B$ = 10000 G, $2 r_c \sim$ 500 nm, thus  chiral edge states do not contribute to the Andreev bound states localized at the edge. We note that this 2$r_c$ value is much larger than $l_e$ = 30 nm of $L$ = 500 nm Gr/WS$_2$ junction. At higher fields, $r_c$ becomes smaller than $l_e$, so that the Landau localization of bulk states and the formation of chiral edge states may become relevant. We observe superconducting pockets even at 70000 G for Gr/WS$_2$ junction for $L$ = 500 nm, and $\delta I_c \sim$ 30 nA \cite{wakamura3}. However, this value is more than one order of magnitude larger than $\delta I_c$ reported in the QH regime with a comparable $r_c$ for shorter and narrower junctions with better quality graphene \cite{Amet}. This may indicate that the supercurrent enhancement by the SOIs can also be effective in the QH regime.  

\section{Discussions and future possibilities}
In this review we showed that strong SOIs can be induced in graphene through heterostructures with TMDs. The susceptibility of graphene's exposed 2D electron gas facilitates the modulation of its transport properties via proximity effects. Owing to interactions between $\pi$ electrons in graphene and $p$- or $d$-orbitals in TMDs, originating from calchogen and heavy atom, respectively, the strong intrinsic SOI in TMDs are transferred to graphene, thereby WAL signals are observed in magneto-transport measurements at low temperatures. While all of the TMDs have stronger SOIs than graphene, monolayer tungsten-based TMDs (WS$_2$ and WSe$_2$) bring considerable enhancement of SOIs in graphene compared with their bulk counterparts or other TMDs.  

In addition to the normal-state transport properties, supercurrent transport through graphene exhibits dramatic changes driven by the strong SOIs. Even for diffusive graphene with relatively low mobility, signatures of robust supercurrent are observed even around 1 T in Gr/WS$_2$ Josephson junctions. Specifically, for the $L$=500 nm Gr/WS$_2$ junctions, dips of the differential resistance $dV/dI$ are observed in magnetic fields as high as 7 T, a signature of superconductivity induced in graphene. These exotic phemena may be driven by the localization of supercurrent near the edges protected by the strong SOIs.

While it is clear that graphene/TMD hybrid structures are an intriguing system to explore, several issues still remain to be clarified. This system has been investigated both theoretically and experimentally, but there are discrepancies in the estimates of the induced SOIs in graphene. Early theoretical reports based on the DFT calculations predict SOI amplitudes of the order of  $\sim$ 1 meV \cite{gmitra2, gmitra3}, smaller than the values obtained in experiments \cite{wakamura1, wakamura2, avsar, zihlmann, eroms1, eroms2, wang1, wang2, Yang1, Yang2}. More recently, the angular dependence of the SOIs induced in graphene by the TMDs has been considered \cite{koshino, david}.  These studies take into account a finite angle between graphene and a TMD layer, more appropriate to real experimental devices. They demonstrate that the amplitudes of the SOIs induced in graphene dramatically change depending on the relative angle between the two layers, and at certain angles the SOIs become a few tens of meV, much larger than the values estimated based on the DFT calculations where the relative angle between the two layers is always set to zero. The discrepancies between the theoretical estimates and experimental ones may also be due to differences in the position of the Fermi level in the band structure of the system. For graphene/TMD bilayers, the band structure is generically the superposition of the band structure of graphene and the TMD. However, the position of the Dirac cone (note that now the linear band crossing is gapped due to the SOIs and bands are spin-split) depends on the combination of graphene and the TMD. It is shown that the amplitudes of the induced SOIs are highly affected by the position of the Fermi level in the band structure \cite{david}. The band structure of graphene/TMD hybrids has been experimentally investigated by angular-resolved spectroscopy (ARPES) measurements \cite{pierucci, henck}. In the case of graphene/MoS$_2$ heterostructures, while the ARPES data show that the Fermi level is located around the middle of the conduction band, the DFT calculations propose that the Fermi level is near the conduction band edge of MoS$_2$ \cite{gmitra2, pierucci}. This discrepancy between the experimental results and calculations may also be one of the reasons for the difference between the theoretical and experimental estimates of the SOIs. To clarify these points further systematic experimental studies are essential including different experimental probes. One of the promising candidates among them is resistively-detected electron-spin resonance (RD ESR) technique, already used to determine the amplitude of the intrinsic SOI in graphene \cite{sichau, sharma}. By using this technique it may be possible to reveal the band structure of graphene/TMD hybrids, thereby more precise values of SOIs can be extracted.     

We now briefly discuss future directions of the graphene/TMD hybrid system. As for normal state transport phenomena, since it is possible to generate spin currents even without ferromagnetic contacts by using nonmagnetic TMDs such as MoS$_2$ or WSe$_2$, it is of great interest to fabricate a device with pristine graphene connected to graphene/TMD electrodes. These electrodes act as a spin current generator and detector. So far there have been reports on spin transport in a graphene/TMD bilayer \cite{vanwees1, valenzuera1, vanwees2, valenzuera2}. However, ferromagnetic metals are exploited as a spin injector or detector in these studies. By using a graphene/TMD bilayer as a spin injector and detector for ''pristine'' graphene instead of conventional ferromagnetic electrodes, it becomes much easier to overcome spin impedance mismatch problem between spin injector/detector and graphene. By depositing TMD flakes partially on graphene, it is realizable to make such a device. 

It is also attractive to explore supercurrent properties in graphene/TMD Josephson junctions. To demonstrate more explicitly the localization of supercurrent, scanning tunneling microscope (STM) may be useful to show the edge states which carry supercurrent. Since supercurrent can flow for as long as 500 nm in graphene/WS$_2$ Josephson junctions, it is also possible to fabricate more complex structures to carry out further investigations of supercurrent transport.   
 
\section{Acknowledgement}
Our studies presented here were realized in collaboration with C. Mattevi, M. Och, P. Palczynski, F. Reale, T. Taniguchi, K. Watanabe, A. D. Chepelianskii, M. Ferrier, N. J. Wu, M. Q. Zhao, A. T. C. Johnson and A. Ouerghi. We sincerely acknowledge strong supports by our cowokers. Our projects are finantially supported in part by the Marie Sklodowska Curie Individual Fellowships (H2020-MSCAIF-2014-659420); the ANR Grants DIRACFORMAG (ANR-14-CE32-003), MAGMA(ANR-16-CE29-0027-02), and JETS (ANR-16-CE30-0029-01); the Overseas Research Fellowships by the Japan Society for the Promotion of Science (2017-684); and the CNRS. K. W. and T. T. acknowledge support from the Elemental Strategy Initiative conducted by the MEXT, Japan, Grant No. JPMXP0112101001, JSPS KAKENHI Grant No. JP20H00354, and the CREST (JPMJCR15F3), JST.
\printcredits

%\bibliographystyle{cas-model2-names}
%\bibliographystyle{crunsrt}
%\bibliography{cas-refs2}
%\bibliography{samplebib}


\end{document}